\documentclass{article}
\usepackage{authblk}
\usepackage{amsmath,amsfonts,amssymb}
\usepackage{txfonts}
\usepackage[T1]{fontenc}
\usepackage{hyperref}

\def\PL#1{Phys.\ Lett.\ {\bf#1}}

\def\PR#1{Phys.\ Rev.\ {\bf#1}}

\def\JoP#1{J.\ Phys.\ {\bf#1}} \def\IJMP#1{Int.\ J. Mod.\ Phys.\ {\bf #1}}

\def\bd{\begin{displaymath}}\def\ed{\end{displaymath}}
\def\be{\begin{equation}}\def\ee{\end{equation}}
\def\bea{\begin{eqnarray}}\def\eea{\end{eqnarray}}
\def\ba{\begin{array}}\def\ea{\end{array}}

\newcommand{\arxiv}[1]{\href{https://arxiv.org/abs/#1}{arXiv:#1}}

\newcommand{\bibxp}[4]{#1, #2, #3, \arxiv{#4}}

\begin{document}
\begin{titlepage}
\title{On interpolations between Jordanian twists}
\author[1]{D. Meljanac{\footnote{Daniel.Meljanac@irb.hr}}}
\author[2]{S. Meljanac{\footnote{meljanac@irb.hr}}}
\author[3]{Z. \v{S}koda{\footnote{zskoda@unizd.hr}}}
\author[4]{R. \v Strajn{\footnote{rina.strajn@unidu.hr}}}
\affil[1]{Division of Materials Physics, Ru\dj er Bo\v skovi\'c Institute, Bijeni\v cka cesta 54, 10002 Zagreb, Croatia}
\affil[2]{Division of Theoretical Physics, Ru\dj er Bo\v skovi\'c Institute, Bijeni\v cka cesta 54, 10002 Zagreb, Croatia}
\affil[3]{
Department of Teachers' Education, University of Zadar, Franje Tudjmana 24, 23000 Zadar, Croatia}
\affil[4]{Department of Electrical Engineering and Computing, University of Dubrovnik, \'{C}ira Cari\'{c}a 4, 20000 Dubrovnik, Croatia}
\date{}
\renewcommand\Affilfont{\itshape\small}
\clearpage\maketitle
\thispagestyle{empty}

\begin{abstract}
We consider two families of Drinfeld twists generated from a simple Jordanian twist further twisted with 1-cochains. Using combinatorial identities, they are presented as a series expansion in the dilatation and momentum generators. These twists interpolate between two simple Jordanian twists. For an expansion of a family of twists $\mathcal{F}_{L,u}$, we also show directly that the 2-cocycle condition reduces to previously proven identities.
\end{abstract}
\end{titlepage}

\section{Introduction}

In order to resolve fundamental problems at the Planck scale, the notion of noncommutative (NC) spacetime was proposed, e.g., the $\kappa$-Minkowski spacetime \cite{LukTol,LukRuegg}. The parameter $\kappa$ is interpreted as the Planck mass or the quantum gravity scale. NC spaces lead to deformed relativistic spacetime symmetries and corresponding dispersion relations. The related symmetries exhibit a Hopf algebra structure. For Hopf algebras there exists a twisting procedure, due to Drinfeld \cite{drinfeld,majid}, providing a laboratory for the deformation of spacetime, its relativistic symmetry and geometric and physical structures. Drinfeld 2-cocycles can be further modified with a 1-cochain \cite{drinfeld,majid}. This can be interpreted as a gauge transformation of the 2-cocycle.

In the 1980's, deformations of $R$-matrices and the related quantum groups, called Jordanian $R$-matrices and Jordanian deformations, have been constructed. The corresponding Drinfeld twist has been written in~\cite{Og}. Furthermore, $r$-symmetric versions of the Jordanian twist, where $r$ is the classical $r$-matrix, were proposed \cite{Tolstoy,GZ}. The Jordanian twist reappeared in the context of the $\kappa$-Minkowski NC spacetime \cite{BP,BuKim}. A relation with the symmetry of the $\kappa$-Minkowski spacetime is established with the introduction of the generators of relativistic symmetries, dilatation $D$ and momenta $p_\alpha$, satisfying $[D,p_\alpha]=-p_\alpha$. Dilatation $D$ is included in the minimal extension of the relativistic spacetime symmetry, the so-called Poincar\'e-Weyl algebra, generated by $\{ M_{\mu\nu},p_\mu,D\}$, where $M_{\mu\nu}$ denote the Lorentz generators. One parameter interpolations between Jordanian twists, generated from a simple Jordanian twist $\mathcal{F}_0$, by twisting with 1-cochains, were studied in \cite{mmmsprd,remarks,cobtw,MMss,MMss2}.

For physical applications in the perturbative approach, it is important to have an expansion in $1/\kappa$. Here, we present two special families of twists induced with 1-cochains and their series expansion in the dilatation and momentum generators. We prove that these twists interpolate between two simple Jordanian twists.

The plan of the paper is as follows. In Section~\ref{sec:FLu} we first recall the definition of twist $\mathcal{F}_{L,u}$~\cite{remarks,cobtw} as a product of 3 exponential expressions and then we derive a new explicit expansion~(\ref{FLu22}) for $\mathcal{F}_{L,u}$. In Section~\ref{sec:coc} we prove that this expansion indeed satisfies the 2-cocycle condition. In Section~\ref{sec:FRu} we discuss an explicit expansion~(\ref{FRuexp}) of another family of Drinfeld twists, $\mathcal{F}^{-1}_{R,u}$.

\section{Family $\mathcal{F}_{L,u}$ of Jordanian twists}
\label{sec:FLu}

In \cite{remarks,cobtw}, the following family of Drinfeld twists was considered
\begin{eqnarray}
\mathcal{F}_{L,u}&=&\text{exp}\Big( -u\left( DA\otimes 1+1 \otimes DA\right) \Big)\, \text{exp}\Big( -\text{ln}(1+A)\otimes D\Big)\, \text{exp}\Big(\Delta (uDA)\Big)\\
&=& \text{exp}\left( \frac{u}{\kappa}\left( DP\otimes 1+1 \otimes DP\right) \right)\, \text{exp}\left( -\text{ln}\left(1-\frac{1}{\kappa}P\right)\otimes D\right) \, \text{exp}\left(-\Delta \left(\frac{u}{\kappa} DP\right)\right)\label{FLu2}
\end{eqnarray}
where generators dilatation $D$ and momenta $P$ satisfy $[P,D]=P$ and $A=-\frac{1}{\kappa}P$. The deformation parameter $\kappa$ is of the order of the Planck mass, $u$ is a real dimensionless parameter and $\Delta$ is the undeformed coproduct. These twists are constructed using a 1-cochain $\omega_L=\text{exp}\left(-\frac{u}{\kappa}DP\right)$ and they satisfy the normalization and cocycle condition. The corresponding deformed Hopf algebra is given by
\begin{eqnarray}
&& \Delta^{\mathcal{F}_{L,u}} (p_\mu) =\mathcal{F}_{L,u}\, \Delta p_\mu\, \mathcal{F}_{L,u}^{-1} =\frac{p_\mu \otimes \left(1+\frac{u}{\kappa}P\right) +\left(1-\frac{1-u}{\kappa}P\right) \otimes p_\mu}{1\otimes 1+u(1-u)\frac{1}{\kappa^2}P\otimes P}\\
&& \Delta^{\mathcal{F}_{L,u}} (D)= \mathcal{F}_{L,u}\,\Delta D\, \mathcal{F}_{L,u}^{-1} = \left( D\otimes \frac{1}{1+\frac{u}{\kappa}P} +\frac{1}{1-\frac{1-u}{\kappa}P}\otimes D\right) \left( 1\otimes 1 +u(1-u)\frac{1}{\kappa^2}P\otimes P\right)\nonumber\\
&& \\
&& S^{\mathcal{F}_{L,u}}(p_\mu)=\frac{p_\mu }{1-(1-2u)\frac{1}{\kappa}P}\\
&& S^{\mathcal{F}_{L,u}}(D)=-\left( \frac{1-(1-2u)\frac{P}{\kappa}}{1+\frac{u}{\kappa}P}\right) D\left(1+\frac{u}{\kappa}P\right) ,
\end{eqnarray}
where $p_\mu,\, \mu\in \{0,1,...,n-1\}$ are momenta in the Minkowski spacetime and $P=v^\alpha p_\alpha$, where $v_\alpha v^\alpha\in \{1,0,-1\}$. For $u=0$, $\mathcal{F}_{L,u=0}$ reduces to the Jordanian twist $\mathcal{F}_{0}=\text{exp}(-\text{ln}(1-\frac{1}{\kappa}P)\otimes D)$. For $u=1/2$, $\mathcal{F}_{L,u=1/2}$ corresponds to the twist proposed in \cite{Tolstoy}. For $u=1$ it was claimed in \cite{remarks} that $\mathcal{F}_{L,u=1}$ is identical to the Jordanian twist $\mathcal{F}_1=\text{exp}(-D\otimes \text{ln}(1+\frac{1}{\kappa}P))$ and this was checked up to the third order in $1/\kappa$.

Twists $\mathcal{F}_{L,u=1}$ and $\mathcal{F}_{1}$ generate the same *-product of arbitrary functions $f(x)$ and $g(x)$. However, these two twists could in principle differ in higher orders of $1/\kappa$. For this reason, we here prove that they are identically equal in all orders in $1/\kappa$, $\mathcal{F}_{L,u=1}=\mathcal{F}_1=\text{exp}\left(-D\otimes \ln (1+P/\kappa)\right)$.

Let us expand the exponential factors
\begin{eqnarray}
&& \text{e}^{\frac{u}{\kappa}(DP\otimes 1+1\otimes DP)} =\sum_{k_1,l_1=0}^\infty \left(\frac{u}{\kappa}\right)^{k_1+l_1}P^{k_1} \binom{D-1}{k_1} \otimes P^{l_1} \binom{D-1}{l_1} \\
&& \text{e}^{-\text{ln}\left(1-\frac{1}{\kappa}P\right)\otimes D} = \sum_{k'=0}^\infty \left(-\frac{1}{\kappa}P\right)^{k'}\otimes \binom{-D}{k'}\\
&& \text{e}^{-\Delta(\frac{u}{\kappa} DP)} =\sum_{k_2,l_2=0}^\infty \left(-\frac{u}{\kappa}\right)^{k_2+l_2} \left( P^{k_2} \otimes P^{l_2}\right) \binom{D\otimes 1+1\otimes D-1\otimes 1}{k_2+l_2} \binom{k_2+l_2}{k_2}.
\end{eqnarray}
Here we use the binomial symbol abbreviation 
\begin{equation}
\binom{T}{k} = \frac{T(T-1)\cdots(T-k+1)}{k!}
\end{equation}
for arbitrary variable $T$ (not necessarily a number) and any positive integer $k$. Then,
\begin{eqnarray}
\mathcal{F}_{L,u} &=& \sum_{k_1,l_1,k',k_2,l_2=0}^\infty (-u)^{k_1+l_1} u^{k_2+l_2} \Bigg( \left(-\frac{P}{\kappa}\right)^{k_1+k'+k_2} \binom{D-1-k'-k_2}{k_1} \otimes \nonumber \\
& \quad & \left(-\frac{P}{\kappa}\right)^{l_1+l_2} \binom{D-1-l_2}{l_1} \binom{-D+l_2}{k'} \Bigg) \binom{D\otimes 1+1\otimes D-1\otimes 1}{k_2+l_2} \binom{k_2+l_2}{k_2}
\end{eqnarray}
Note that from $[P,D]= P$ it follows that $DP=P(D-1)$, and generally
\begin{eqnarray}
&& f(D)P= Pf(D-1) \\
&& f(D)P^m= P^m f(D-m).
\end{eqnarray}
Next we define $k=k_1+k'+k_2$, $l=l_1+l_2$. Then,
\begin{eqnarray}
\mathcal{F}_{L,u} &=& \sum_{k,l=0}^\infty \left( \left(-\frac{P}{\kappa}\right)^k\otimes \left(-\frac{P}{\kappa}\right)^l\right) \sum_{k'=0}^k u^{k-k'+l} \sum_{k_1,l_1,k_2,l_2\geq 0; \,k_1+k_2=k-k'; \,l_1+l_2=l} (-1)^{k_1+l_1} \Bigg( \binom{D-1-k+k_1}{k_1} \otimes \nonumber \\
&\quad & \binom{D-1-l+l_1}{l_1}\binom{-D+l-l_1}{k'} \Bigg) \binom{D\otimes 1+1\otimes D-1\otimes 1}{k_2+l_2} \binom{k_2+l_2}{k_2}.\label{FLu13}
\end{eqnarray}
Let us define $x=D\otimes 1, \, y=1\otimes D$, satisfying $[x,y]=0$. Using the following sequence of identities
\begin{eqnarray}
&& \binom{y-1-l_2}{l_1} = (-1)^{l_1} \binom{-y+l_2+l_1}{l_1} = (-1)^{l_1} \binom{-y+l}{l_1}\label{id1}\\
&& \binom{x+y-1}{k_2+l_2} =(-1)^{k_2+l_2} \binom{-x-y+k_2+l_2}{k_2+l_2}\\
&& \binom{-x-y+k_2+l_2}{k_2+l_2} \binom{k_2+l_2}{k_2} =\binom{-x-y+k_2+l_2}{k-k'-k_1} \binom{-x-y+l_2}{l_2} \\
&& \sum_{k_1,k_2\geq 0; \, k_1+k_2=k-k'} \binom{x-1-k+k_1}{k_1} \binom{-x-y+k_2+l_2}{k-k'-k_1} = \binom{-y-k'+l_2}{k-k'}\nonumber \\
&& \quad = \sum_{k_1,k_2\geq 0; \, k_1+k_2=k-k'} (-1)^{k_1} \binom{-x+k}{k_1}\binom{-x-y+k_2+l_2}{k-k'-k_1} \\
&& \binom{-y+l}{l_1} \binom{-y+l_2}{k'} = \binom{-y+l}{k'}\binom{-y+l-k'}{l_1} \\
&& \binom{-y-k'+l_2}{k-k'}\binom{-y+l-k'}{l_1}=\binom{-y+l-k'}{k-k'}\binom{-y+l-k}{l_1}\\
&& \sum_{l_1,l_2\geq 0;\, l_1+l_2=l} (-1)^{l_1} \binom{-y+l-k}{l_1} \binom{-x-y+l_2}{l_2} =\binom{-x+k}{l}\\
&& \binom{-y+l-k'}{k-k'} \binom{-y+l}{k'} =\binom{k}{k'} \binom{-y+l}{k},\label{idz}
\end{eqnarray}
and using $k=k_1+k_2+k'$ and $l=l_1+l_2$, we get
\begin{eqnarray}
&& \binom{y-1-l_2}{l_1} =(-1)^{l_1} \binom{-y+l}{l_1}\\
&& \binom{x+y-1}{k_2+l_2}= (-1)^{k_2+l_2} \binom{-x-y+k_2+l_2}{k_2+l_2}\\
&& \binom{-x-y+k_2+l_2}{k_2+l_2} \binom{k_2+l_2}{k_2} =\binom{-x-y+k_2+l_2}{k_2} \binom{-x-y+l_2}{l_2}\\
&& \sum_{k_1,k_2\geq 0;\, k_1+k_2=k-k'} \binom{x-1-k+k_1}{k_1} \binom{-x-y+k_2+l_2}{k_2} = \sum_{k_1,k_2 \geq 0;\, k_1+k_2=k-k'} (-1)^{k_1}\nonumber\\
&& \\
&& \binom{-x+k}{k_1}\binom{-x-y+k_2+l_2}{k_2} =\binom{-y-k'+l_2}{k-k'}\\
&& \binom{-y+l}{l_1}\binom{-y+l_2}{k'} = \binom{-y+l}{k'} \binom{-y+l-k'}{l_1}\\
&& \binom{-y-k'+l_2}{k-k'}\binom{-y+l-k'}{l_1} = \binom{-y+l-k'}{k-k'} \binom{-y+l-k}{l_1}\\
&& \sum_{l_1,l_2\geq 0;\, l_1+l_2=l} (-1)^{l_1}\binom{-y+l-k}{l_1} \binom{-x-y+l_2}{l_2} =\binom{-x+k}{l}\\
&& \binom{-y+l-k'}{k-k'} \binom{-y+l}{k'} =\binom{k}{k'} \binom{-y+l}{k}.
\end{eqnarray}
After performing a summation over $k'$, we finally obtain 
\begin{equation}\label{FLu22}
\mathcal{F}_{L,u}= \sum_{k,l=0}^\infty \left(\frac{1}{\kappa}\right)^{k+l} \binom{-D}{l} (u-1)^k P^k \otimes \binom{-D}{k} (uP)^l.
\end{equation}
Similarly, one can obtain the expression for $\mathcal{F}_{L,u}^{-1}$,
\begin{equation}
\mathcal{F}_{L,u}^{-1} =\sum_{k,l=0}^\infty (u-1)^k \left(\frac{P}{\kappa}\right)^k \otimes \left(u\frac{P}{\kappa}\right)^l \Bigg( \binom{D-1}{l} \otimes \binom{D-1}{k} +\binom{D-1}{l} \otimes \binom{D-1}{k-1}+ \binom{D-1}{l_1} \otimes \binom{D-1}{k} \Bigg).
\end{equation}
For $u=0$,
\begin{equation}
\mathcal{F}_{L,u=0} =\sum_{k=0}^\infty \left(-\frac{P}{\kappa}\right)^k \otimes \binom{-D}{k} = \text{exp}\left(-\text{ln}\left(1-\frac{1}{\kappa}P\right)\otimes D\right) =\mathcal{F}_0.
\end{equation}
For $u=1$,
\begin{equation}
\mathcal{F}_{L,u=1} = \sum_{l=0}^\infty \binom{-D}{l} \otimes \left(\frac{P}{\kappa}\right)^l =\text{exp}\left(-D\otimes \text{ln}\left(1+\frac{1}{\kappa}\right)\right) =\mathcal{F}_1.
\end{equation}
Hence, the family of twists $\mathcal{F}_{L,u}$ interpolates between twists $\mathcal{F}_0$ and $\mathcal{F}_1$.

One can check that differentiation of $\mathcal{F}_{L,u}$ \eqref{FLu2} and $\mathcal{F}_{L,u}$ \eqref{FLu22} with respect to parameter $u$ gives the same differential equation. The expressions \eqref{FLu2} and \eqref{FLu22} are furthermore equal at the initial 
value $u = 0$. This confirms that our result in \eqref{FLu22} is correct.

\section{2-cocycle condition for twist $\mathcal{F}_{L,u}$ from~(\ref{FLu22})}
\label{sec:coc}

In this section, we present a direct proof that $\mathcal{F}_{L,u}$, defined by the power expansion~(\ref{FLu22}), satisfies the 2-cocycle condition,
\begin{equation}\label{eq:Fcoc}
  (\mathcal{F}_{L,u}\otimes 1)(\Delta\otimes\mathrm{id})\mathcal{F}_{L,u}
  = (1\otimes\mathcal{F}_{L,u})(\mathrm{id}\otimes\Delta)\mathcal{F}_{L,u}.
\end{equation}
The proof is pretty analogous to the proof of the 2-cocycle condition for the twist $\mathcal{F}_{GZ,u}^{-1}$ in~\cite{MMss}.
From the expansion~(\ref{FLu22}) for $\mathcal{F}_{L,u}$,
we introduce an auxiliary expression 
$f_n$, $n = 0,1,2,3,\ldots,$ not depending on parameter $\kappa$, such that 
\begin{equation}\label{eq:FLufn}
  \mathcal{F}_{L,u}= \sum_{n=0}^\infty \left(\frac{1}{\kappa}\right)^{n} f_n
\end{equation}
Comparing to~(\ref{FLu22}), we obtain $f_0 = 1\otimes 1$ and for $n>1$
\begin{equation}\label{eq:fn}
f_n = \sum_{k+l = n} \binom{-D}{l} (u-1)^k P^k \otimes \binom{-D}{k} (uP)^l.
\end{equation}
Inserting~(\ref{eq:FLufn}) into the 2-cocycle condition~(\ref{eq:Fcoc}) gives an equivalent condition in terms of $f_i$-s, namely, for all $n\geq 0$, 
\begin{equation}\label{eq:cocfi}
  \sum_{i = 0}^n (f_{n-i}\otimes 1)(\Delta\otimes{\mathrm{id}})f_i
  = \sum_{i = 0}^n(1\otimes f_{n-i})(\mathrm{id}\otimes\Delta)f_i.
\end{equation}
Once we substitute $f_i$ from~(\ref{eq:fn}) into the condition~(\ref{eq:cocfi}) we obtain an equality of two polynomials in $u$. 
Now notice that, for fixed $n$, the polynomials $(u-1)^ku^{n-k}$ for $0\leq k\leq n$ are linearly independent. Indeed, the condition $\sum_{k=0}^n c_k (u-1)^k u^{n-k} = 0$ for some $c_k$ can be written in the basis $u^n,u^{n-1},\ldots,u,1$ as the system $(c_0,\ldots, c_n)\, A = 0$, where $A$ is an $n\times n$ {\em triangular} matrix with rows $(1,0,0,\ldots,0)$, $(1,-1,0,\dots,0)$, $(1,-2,1,0,\ldots 0)$, $(1,-3,3,-1,0,\ldots,0),\ldots$, hence with $\pm 1$ on the diagonal. In particular, the determinant of this matrix is $(-1)^n\neq 0$, hence $c_i = 0$. We conclude that the polynomials in $u$ on two sides of Equation~(\ref{eq:cocfi}) are equal iff the corresponding coefficients in front of $(u-1)^k u^{n-k}$ for each $k$ are equal. Power $k$ in $(u-1)^k$ is obtained from the powers $(u-1)^{k_1}$ in $f_{n-i}$ and $u^{k-k_1}$ in $f_i$ for all $0\leq k_1\leq k$ and, similarly, power $l$ in $u^l$ is obtained from $u^{l_1}$ in $f_{n-i}$ and $u^{l-l_1}$ in $f_i$ for all $0\leq l_1\leq l$. For fixed $k$ and $l = n - k$, this yields the condition
\begin{eqnarray*}
  \sum_{k_1 = 0}^k \sum_{l_1 = 0}^l \left[\binom{-D}{l_1} P^{k_1} \otimes \binom{-D}{k_1} P^{l_1}\otimes 1\right]\left[\Delta\left(\binom{-D}{l-l_1}P^{k-k_1}\right)\otimes\binom{-D}{k-k_1}P^{l-l_1}\right] 
  \\
  =
  \sum_{k_1 = 0}^k \sum_{l_1 = 0}^l\left[1\otimes\binom{-D}{l_1}P^{k_1}\otimes
    \binom{-D}{k_1}P^{l_1}\right]\left[\binom{-D}{l-l_1}P^{k-k_1}\otimes
    \Delta\left(\binom{-D}{k-k_1}P^{l-l_1}\right)\right].
\end{eqnarray*}
Now we want to achieve an ordering in the algebra generated by $P,D$ with $[P,D] = P$ such that all $P$-s are to the right. For reordering we use the identity (which can be easily obtained by induction) that
\begin{equation}
\binom{D-r}{m} P^r = P^r \binom{D}{m}.
\end{equation}
We use $\Delta(D) = D\otimes 1 + 1\otimes D$, obtaining 
\begin{eqnarray*}
  \sum_{k_1 = 0}^k \sum_{l_1 = 0}^l \left[\left(\binom{-D}{l_1}\otimes\binom{-D}{k_1}\right)\binom{(-D-k_1)\otimes 1 + 1\otimes (-D-l_1)}{l-l_1}\otimes 1\right]
  \left[(P^{k_1}\otimes P^{l_1})\Delta(P^{k-k_1})\otimes\binom{-D}{k-k_1}P^{l-l_1}\right] 
  \\
  =
  \sum_{k_1 = 0}^k \sum_{l_1 = 0}^l\left[1\otimes\left(\binom{-D}{l_1}\otimes\binom{-D}{k_1}\right)\binom{(-D-k_1)\otimes 1 + 1\otimes(-D-l_1)}{k-k_1}\right]
  \left[\binom{-D}{l-l_1}P^{k-k_1}\otimes(P^{k_1}\otimes P^{l_1})\Delta(P^{l-l_1})\right].
\end{eqnarray*}
In this equation, we evaluate
$$\Delta(P^{k-k_1}) = \sum_{a = 0}^{k-k_1}\binom{k-k_1}{a}\,P^{k-k_1-a}\otimes P^a,$$ 
$$\Delta(P^{l-l_1}) = \sum_{b = 0}^{l-l_1}\binom{l-l_1}{b}\,P^{b}\otimes P^{l-l_1-b},$$
and define shortcuts 
$$
x = - D\otimes 1\otimes 1,\,\,\,\,\,
y = - 1\otimes D\otimes 1,\,\,\,\,\,
z = - 1\otimes 1\otimes D,
$$
which behave as independent mutually commuting variables, however they do not commute with the tensor products involving $P$. Adding a number to $x,y$ or $z$ will, in this notation, mean adding this number times $1\otimes 1\otimes 1$. Then we can write simply
\begin{eqnarray*}
  \sum_{k_1=0}^k\sum_{l_1=0}^l\sum_{a = 0}^{k-k_1}\binom{x}{l_1}\binom{y}{k_1}\binom{x-k_1+y-l_1}{l-l_1}\binom{k-k_1}{a}\binom{z}{k-k_1}\left(P^{k-a}\otimes P^{l_1+a}\otimes P^{l-l_1}\right)=
  \\
  = \sum_{k_1=0}^k\sum_{l_1=0}^l\sum_{b = 0}^{l-l_1}\binom{y}{l_1}\binom{z}{k_1}
  \binom{y-k_1+z-l_1}{k-k_1}\binom{x}{l-l_1}\binom{l-l_1}{b}
  \left(P^{k-k_1}\otimes P^{k_1+b}\otimes P^{l-b}\right).
\end{eqnarray*}
Regarding that we use the normal ordering (or in other words that we split the $P$-part and $D$-part of the algebra into a semidirect product), we should compare the coefficients in front of $P^A\otimes P^B\otimes P^C$ for each triple $A,B,C$. Once $A$ and $C$ are fixed one can see that $B = k+l - A - C$ holds for each summand.  The left-hand side has $l_1$ fixed by $l-l_1 = C$ and the right-hand side has $k_1$ fixed by taking into account $k-k_1 = A$. In the left-hand side sum, $k-a = A$ hence appearance of the binomial coefficient $\binom{k-k_1}{k-A}$ means that only the terms with $k_1\leq A\leq k$ survive, while in the right-hand side sum only the terms with $l_1\leq C\leq l$ survive. We also make all necessary replacements, expressing other quantities in terms of $A,C,k,l$. This reduces
the 2-cocycle condition to the identities
\begin{eqnarray}
  \binom{x}{l-C}\sum_{k_1 = 0}^A \binom{y}{k_1}\binom{x+y-k_1+C-l}{C}\binom{k-k_1}{k-A}\binom{z}{k-k_1}\nonumber \\
  = \binom{z}{k-A}\sum_{l_1=0}^C \binom{x}{l-l_1}\binom{y}{l_1}
  \binom{y+z-l_1+A-k}{A}
  \binom{l-l_1}{l-C}.
  \label{eq:bigidentn}
\end{eqnarray}
These identities differ from the Equation (2.3) in~\cite{MMss}, which is proven in the Appendix~A of~\cite{MMss}, only by a simple interchange of summation indices $k_1\mapsto k-k_1$ for the first sum and $l_1\mapsto l-l_1$ for the second sum, with correct bounds of summation. Hence the identities~(\ref{eq:bigidentn}) and therefore also the 2-cocycle condition~(\ref{eq:Fcoc}) hold.

\section{Family $\mathcal{F}_{R,u}$ of Jordanian twists}
\label{sec:FRu}

Another family of twists induced with a 1-cochain $\omega_R=\text{exp}\left(-\frac{u}{\kappa}PD\right)$ is \cite{cobtw} 
\begin{equation}\label{FRu}
\mathcal{F}_{R,u}=\text{e}^{\frac{u}{\kappa}(PD\otimes 1+1\otimes PD)}\, \text{e}^{-\text{ln}\left(1-\frac{1}{\kappa}P\right)\otimes D}\, \text{e}^{-\Delta\left(\frac{u}{\kappa} PD\right)}.
\end{equation}
These twists satisfy the normalization and cocycle condition. The corresponding deformed Hopf algebra is given by
\begin{eqnarray}
&& \Delta^{\mathcal{F}_{R,u}}(p_\mu) =\mathcal{F}_{R,u}\, \Delta p_\mu\, \mathcal{F}_{R,u}^{-1} =\frac{p_\mu \otimes \left( 1+\frac{u}{\kappa}P\right) +\left( 1-\frac{1-u}{\kappa}P\right) \otimes p_\mu}{1\otimes 1+\frac{u(1-u)}{\kappa^2}P\otimes P} \\
&& \Delta^{\mathcal{F}_{R,u}} (D)=\mathcal{F}_{R,u}\, \Delta D\, \mathcal{F}_{R,u}^{-1}=\left( 1\otimes 1+\frac{u(1-u)}{\kappa^2} P\otimes P\right) \left( D\otimes \frac{1}{1+\frac{u}{\kappa}P} +\frac{1}{1-\frac{1-u}{\kappa} P} \right)\nonumber \\
  && \\
&& S^{\mathcal{F}_{R,u}} (p_\mu) =-\frac{p_\mu}{1-\frac{1-2u}{\kappa}P} \\
&& S^{\mathcal{F}_{R,u}} (D)= -\left( 1-(1-u)\frac{P}{\kappa}\right) D \left(\frac{1-(1-2u)\frac{P}{\kappa}}{1-(1-u)\frac{P}{\kappa}}\right).
\end{eqnarray}
For $u=0$, $\mathcal{F}_{R,u=0}$ reduces to the Jordanian twist $\mathcal{F}_0$ and for $u=1$ we shall prove that $\mathcal{F}_{R,u=1} =\mathcal{F}_1$.

Let us consider the twist
\begin{equation}\label{F-1Ru40}
\mathcal{F}_{R,u}^{-1}=\text{e}^{\Delta\left( \frac{u}{\kappa}P D\right)}\,\text{e}^{\text{ln}\left(1-\frac{1}{\kappa}P\right)\otimes D}\, \text{e}^{-\frac{u}{\kappa} (PD\otimes 1+1\otimes PD)}.
\end{equation}
After expanding exponential factors, we find
\begin{eqnarray}
\mathcal{F}_{R,u}^{-1} &=& \sum_{k,l=0}^\infty \left(\left(-\frac{P}{\kappa}\right)^k\otimes \left(-\frac{P}{\kappa}\right)^l\right) \sum_{k'=0}^k (-u)^{k-k'+l} \sum_{k_1,l_1,k_2,l_2\geq 0;\, k_1+k_2=k-k';\, l_1+l_2=l} (-1)^{k_1+l_1} \times \nonumber \\
&& \left( \binom{D}{k_1} \otimes \binom{D-l_1}{k'}\binom{D}{l_1}\right) \binom{D\otimes 1+1\otimes D-(k'+k_1+l_1)1\otimes 1}{k_2+l_2} \binom{k_2+l_2}{k_2}.
\end{eqnarray}

After using a similar sequence of identities as for $\mathcal{F}_{L,u}$ we get
\begin{eqnarray}
&& \sum_{k_1,k_2,l_1,l_2 \geq 0;\, k_1+k_2=k-k';\, l_1+l_2=l} (-1)^{k_1+l_1} \binom{x}{k_1} \binom{y}{l_1} \binom{y-l_1}{k'} \binom{x+y-(k'+k_1+l_1)}{k_2+l_2} \binom{k_2+l_2}{k_2}\nonumber \\
&& =\binom{k}{k'}\binom{x}{y}\binom{y}{k} \sum_{k_1,l_1=0}^\infty (-1)^{k_1+l_1} \binom{x}{k_1} \binom{y}{l_1} \binom{y-l_1}{k'} \binom{x+y-(k'+k_1+l_1)}{k-k'-k_1+l-l_1} \binom{k_2+l_2}{k_2} \nonumber \\
&& =\sum_{k_1,l_1=0}^\infty (-1)^{k_1+l-1} \binom{x}{k_1}\binom{y}{l_1}\binom{y-l_1}{k'} \frac{(x+y-k'-k_1-l_1)^{\langle k-k'-k_1\rangle} (x+y-k-l_1)^{\langle l-l_1\rangle}}{(k-k'-k_1)!(l-l_1)!} \nonumber \\
&& =\sum_{l_1=0}^\infty (-1)^{l_1} \binom{y}{l_1}\binom{y-l_1}{k'}\binom{y-k'-l_1}{k-k'}\binom{x+y-k-l_1}{l-l_1} \nonumber \\
&& =\sum_{l_1=0}^\infty (-1)^{l_1} \binom{y}{k'}\binom{y-k'}{l_1}\binom{y-k'-l_1}{k-k'}\binom{x+y-k-l_1}{l-l_1} \nonumber \\
&& =\sum_{l_1=0}^\infty (-1)^{l_1} \binom{y}{k'}\binom{y-k'}{k-k'}\binom{y-k}{l_1}\binom{x+y-k-l_1}{l-l_1}\nonumber \\
&& =\binom{x}{l}\binom{y}{k'}\binom{y-k'}{k-k'}=\binom{k}{k'}\binom{x}{l}\binom{y}{k}.
\end{eqnarray}
After performing a summation over $k'$, we get
\begin{equation}\label{FRuexp}
\mathcal{F}_{R,u}^{-1}=\sum_{k,l=0}^\infty (u-1)^k\left(\frac{P}{\kappa}\right)^k\binom{D}{l} \otimes \left(\frac{u}{\kappa}P\right)^l\binom{D}{k}.
\end{equation}
Similarly, for $\mathcal{F}_{R,u}$, one can obtain
\begin{equation}
\mathcal{F}_{R,u} =\sum_{k,l=0}^\infty \left( \binom{-D-1}{l} \otimes \binom{-D-1}{k} +\binom{-D-1}{l}\otimes \binom{-D-1}{k-1} +\binom{-D-1}{l-1}\otimes \binom{-D-1}{k} \right) (u-1)^k \left(\frac{P}{\kappa}\right)^k \otimes \left(\frac{u}{\kappa}P\right)^l.
\end{equation}
For $u=0$,
\begin{equation}
\mathcal{F}_{R,u=0}^{-1}= \sum_{k=0}^\infty \left(-\frac{P}{\kappa}\right)^k \otimes \binom{D}{k} =\text{e}^{\text{ln}(1-\frac{1}{\kappa}P)\otimes D}= \mathcal{F}_0^{-1}.
\end{equation}
For $u=1$,
\begin{equation}
\mathcal{F}_{R,u=1}^{-1}= \sum_{l=0}^\infty \binom{D}{l}\otimes \left(\frac{P}{\kappa}\right)^l =\text{e}^{D\otimes \text{ln}(1+\frac{1}{\kappa}P)}= \mathcal{F}_1^{-1}.
\end{equation}
Hence, the family of twists $\mathcal{F}_{R,u}$ interpolates between two Jordanian twists $\mathcal{F}_0$ and $\mathcal{F}_1$. We point out that for $u=1/2$, $\mathcal{F}_{R,u=1/2}^{-1}=\mathcal{F}_{GZ}^{-1}$, where $\mathcal{F}_{GZ}^{-1}$ is the twist proposed in~\cite{GZ}, theorem 2.20. Hence, the twist $\mathcal{F}_{R,u=1/2}$ is given in~\eqref{FRu} and satisfies the normalization and 2-cocycle condition because it is obtained by gauge transformation from a 2-cocycle.

Note that differentiation of $\mathcal{F}^{-1}_{R,u}$, \eqref{F-1Ru40}, and $\mathcal{F}^{-1}_{R,u}$, \eqref{FRuexp}, with respect to parameter $u$, gives the same differential equation for $\mathcal{F}^{-1}_{R,u}$ in \eqref{F-1Ru40} and \eqref{FRuexp}. The boundary conditions at $u=0$ in \eqref{F-1Ru40} and \eqref{FRuexp} are equal. Hence \eqref{F-1Ru40} and \eqref{FRuexp} are equal for arbitrary $u$ \cite{MMss}. The 2-cocycle condition for $\mathcal{F}^{-1}_{R,u}$ has also been directly proven in~\cite{MMss}.

The relation between $\mathcal{F}_{L,u}$ and $\mathcal{F}_{R,u}$ is given \cite{cobtw}
\begin{equation}
\mathcal{F}^{-1}_{R,u}= \mathcal{F}^{-1}_{L,u} \frac{1}{1\otimes 1 +\frac{u(1-u)}{\kappa^2} P\otimes P}.
\end{equation}
This relation between $\mathcal{F}_{L,u}$ and $\mathcal{F}_{R,u}$ gives an independent check of our explicit calculations in sections 2 and 3.

Note that for $u = 1$, $\mathcal{F}_{L,u=1}= \mathcal{F}_{R,u=1}$. Generally, it holds
\begin{equation}
\text{exp}\left(\frac{1}{\kappa}(DP+vP)\otimes 1+1\otimes \frac{1}{\kappa}(DP+vP)\right)\, \text{exp}\left(-\text{ln}\left(1-\frac{1}{\kappa}P\right)\otimes D\right) \, \text{exp}\left(-\frac{1}{\kappa}\Delta (DP+vP)\right)=\mathcal{F}_1,
\end{equation}
for every $v\in\mathbb{R}$. Hence for different cochains at $u = 1$ we get the same Jordanian twist $\mathcal{F}_1$.

\textbf{Concluding remarks.} From the twists $\mathcal{F}_{L,u}$ and $\mathcal{F}_{R,u}$ one can calculate realizations of NC coordinates $\hat{x}_{\mu}$ in terms of Heisenberg algebra generators \cite{cobtw} (see also \cite{Mercati,nas1608,mmss2}). They correspond to NC coordinates generating the $\kappa$-Minkowski spacetime and satisfy
\begin{equation}
[\hat{x}_{\mu}, \hat{x}_{\nu}] = \frac{i}{\kappa} (v_{\mu} \hat{x}_{\nu} - v_{\nu} \hat{x}_{\mu}).
\end{equation}
Using the methods in \cite{Mercati,nas1608,mmss2,govindarajan}, the corresponding star products are calculated in \cite{cobtw}. Note that the physical interpretation depends on the realizations of the NC coordinates $\hat{x}_{\mu}$ \cite{mmmsprd,Mercati}. Particularly the spectrum of the relativistic hydrogen atom, depending on the parameter $u$, was investigated in \cite{mmmsprd}.

Applications of Drinfeld twists, interpolating between Jordanian twists, to models of field theories and gravity on the $\kappa$-Minkowski spacetime with Poincar\'e-Weyl symmetry will be presented elsewhere, continuing work in~\cite{mmmsprd,govindarajan,geodesic}.

\section*{Acknowledgements}

We thank D.Svrtan for a discussion on combinatorial identities. Z.~\v{S}. has been partly supported by the Croatian Science Foundation under the Project ``New Geometries for Gravity and Spacetime'' (IP-2018-01-7615). This work is registered preprint RBI-ThPhys-2020-37 of Ru\dj er Bo\v{s}kovi\'c Institute.

\footnotesize{

}

\begin{thebibliography}{99}
\bibitem{LukTol} J.~Lukierski, A.~Nowicki, H.~Ruegg, V.~N.~Tolstoy,
$q$-deformation of Poincar\'e algebra, Phys. Lett. B 264, 331 (1991)

\bibitem{LukRuegg} J.~Lukierski, A.~Nowicki,~H. Ruegg, New quantum
Poincar\'e algebra and $\kappa$-deformed field theory, Phys. Lett. B293,
344 (1992)

\bibitem{drinfeld} V.~G.~Drinfel'd, Hopf algebras and the quantum
Yang-Baxter equation, Soviet Math. Dokl. 32, 254 (1985)

\bibitem{majid} S. Majid, Foundations of Quantum Group Theory, Cambridge
University Press

\bibitem{Og} O.~Ogievetsky, Hopf structures on the Borel subalgebra of
$\operatorname{sl}(2)$, in: Proceedings of Winter School in Geometry and Physics, Zdikov, January 1993, Supplemento ai Rendiconti del Circolo Matematico di Palermo, Serie II {\bf 37} (1994) 185.

\bibitem{Tolstoy}
\bibxp{V.~N.~Tolstoy}{Twisted quantum deformations of Lorentz and Poincar\'e algebras}{Bulg. J. Phys. \textbf{35} 441-459 (2008)}{0712.3962}.\\
\bibxp{V.~N.~Tolstoy}{Quantum Deformations of Relativistic Symmetries}{\textit{Invited talk at the XXII Max Born Symposium "Quantum, Super and Twistors", September 27-29, 2006, Wroclaw (Poland), in honour of Jerzy Lukierski}}{0704.0081}.

\bibitem{GZ}
\bibxp{A.~Giaquinto, J.~J.~Zhang}{Bialgebra actions, twists,and universal deformation formulas}{J. Pure Appl. Alg. {\bf 128}, 133 (1998)}{hep-th/9411140}

\bibitem{BP} A.~Borowiec, A.~Pacho\l , $\kappa$-Minkowski spacetime as the
result of Jordanian twist deformation, Phys.Rev. {\bf D 79}:045012 (2009) \href{https://arxiv.org/abs/0812.0576}{arXiv:0812.0576}

\bibitem{BuKim}
J.-G. Bu, J. H. Yee, H.-C. Kim, Differential structure on kappa-Minkowski spacetime realized as module of twisted Weyl algebra, Phys.Lett. {\bf B 679}, 486-490 (2009), \href{https://arxiv.org/abs/0903.0040}{arXiv:0903.0040}

\bibitem{mmmsprd}
 \bibxp{D. Meljanac, S. Meljanac, S. Mignemi, R. \v{S}trajn}{Kappa-deformed phase space, Jordanian twists, Lorentz-Weyl algebra and dispersion relations}{\PR{D99}:12,126012 (2019)}{1903.08679}

\bibitem{remarks}
\bibxp{S.~Meljanac, D.~Meljanac, A.~Pacho\l, D.~Pikuti\'c}{Remarks on simple interpolation between Jordanian twists}{\JoP{A50}, no.26, 265201 (2017)}{1612.07984}.

\bibitem{cobtw} A. Borowiec, S. Meljanac, D. Meljanac, A. Pacho\l, Interpolations between Jordanian twists induced by coboundary twists, SIGMA {\bf 15} (2019) 054 \href{https://arxiv.org/abs/1812.05535}{arXiv:1812.05535}.

\bibitem{MMss}
D.~Meljanac, S.~Meljanac, Z. \v{S}koda, R. \v{S}trajn, {One parameter family of Jordanian twists}, SIGMA {\bf 15} (2019) 082, \href{https://arxiv.org/abs/1904.03993}{arXiv:1904.03993}

\bibitem{MMss2}
D. Meljanac, S. Meljanac, Z. \v{S}koda, R. \v{S}trajn, Interpolations between Jordanian twists, the Poincar\'e-Weyl algebra and dispersion relations, Int. J. Mod. Phys. {\bf A 35}:8, 2050034 (2020) 15pp., \href{https://arxiv.org/abs/1911.03967}{arxiv:1911.03967}

\bibitem{Mercati}
\bibxp{S.~Meljanac, D.~Meljanac, F.~Mercati, D.~Pikuti\'{c}}{Noncommutative spaces and Poincar\'{e} symmetry}{\PL{B766} 181--185 (2017)}{1610.06716}

\bibitem{nas1608}
\bibxp{D. Meljanac, S. Meljanac, S. Mignemi, R. \v{S}trajn}{Snyder-type spaces, twisted Poincar\'{e} algebra and addition of momenta}{\IJMP{A32}, 1750172 (2017)}{1608.06207}

\bibitem{mmss2} S.~Meljanac, D.~Meljanac, A.~Samsarov, M.~Stoji\'c, $\kappa$-deformed Snyder spacetime, Mod.\ Phys.\ Lett. {\bf A25}:579-590 (2010) \href{https://arxiv.org/abs/0912.5087}%
{arXiv:0912.5087}

\bibitem{govindarajan} T.~R.~Govindarajan, K.~S.~Gupta, E.~Harikumar,
S.~Meljanac, D.~Meljanac, Twisted statistics in $\kappa$-Minkowski
spacetime, Phys.Rev. {\bf D77}:105010 (2008) \href{https://arxiv.org/abs/0802.1576}%
{arXiv:0802.1576}

\bibitem{geodesic}
  E. Harikumar, T. Juri\'{c}, S. Meljanac, Geodesic equation in $\kappa$-Minkowski spacetime, Phys. Rev. {\bf D 86}, 045002 (2012), \href{https://arxiv.org/abs/1203.1564}{arxiv:1203.1564}

\end{thebibliography}
\end{document}